\documentclass[conference]{IEEEtran}
\ifCLASSINFOpdf
   \usepackage[pdftex]{graphicx}
\else
\fi
\usepackage[cmex10]{amsmath}
\usepackage{amssymb}
\usepackage{dsfont}
\usepackage{algpseudocode}
\usepackage{subfloat}
\usepackage[Algorithm, boxed]{algorithm}

\usepackage{etoolbox}
\makeatletter
\patchcmd{\maketitle}{\@copyrightspace}{}{}{}
\makeatother

\usepackage{caption}
\usepackage{subcaption}

\begin{document}
%
\title{Finding compact communities in large graphs}

\author{\IEEEauthorblockN{Jean Creusefond}
\IEEEauthorblockA{GREYC, Normandy University\\
Caen, France}
\and
\IEEEauthorblockN{Thomas Largillier}
\IEEEauthorblockA{GREYC, Normandy University\\
Caen, France}
\and
\IEEEauthorblockN{Sylvain Peyronnet}
\IEEEauthorblockA{Qwant \& ix-labs\\
Rouen, France}}


%


\maketitle

\begin{abstract}
This article presents an efficient hierarchical clustering algorithm that solves the problem of \textit{core community detection}.
It is a variant of the standard community detection problem in which we are particularly interested in the connected core of communities.
To provide a solution to this problem, we question standard definitions on communities and provide alternatives.
We propose a function called \textit{compactness}, designed to assess the quality of a solution to this problem.
Our algorithm is based on a graph traversal algorithm, the LexDFS.
The time complexity of our method is in $O(n\times log(n))$.
Experiments show that our algorithm creates highly compact clusters.

\end{abstract}


%
\IEEEpeerreviewmaketitle

\section{Introduction}
\label{sec:intro}

Every person that has friends, acquaintances or any social ties is part of a social network.
This network is then used to interact with people.
The more the technology developed, the more these interactions began to leave traces, creating virtual networks.
These networks exist in many forms: web social network, collaboration, communication, \textit{etc}.

Many properties of such networks have been discovered in the last
decades.  The property that will be discussed in this article is that
people naturally form groups, creating dense structures called
communities.  These communities play a central part in the structure
of the network.  Indeed, we naturally classify our relationships
depending on the community they belong (family, college friends,
\textit{etc.}), which implies breaking the network down into small and
meaningful pieces.  Such structures arise in most of the naturally
formed networks.  Two problems arise when trying to find these
communities: defining them, and designing an algorithm matching the
definition.

In social networks, defining communities has practical political and economical consequences.
For instance, when considering someone for targeted marketing, knowing whether or not he is in the targeted community is crucial.
In the same way, the amount of people that is identified as close to a political party by social network analysis may be completely different depending on the way of considering the community detection problem.
The method proposed in this article is designed for social networks.
However, we note that it might also be applicable for general purposes.

Strong and weak communities are objects that were defined by Radicchi \textit{et al.}~\cite{radicchi_defining_2004}, and these definitions are commonly accepted.
A strong community is one whose users have more links inside of the community than outside.
The weak community definition relaxes the strong one by stating that there are more links connecting the inside of a community than connecting the outside.
However, the direct use of these definitions is quickly intractable, since the number of subsets of vertices following these definitions is often exponentially large.
That is why quality functions were developed, to be able to compare communities and to give a solution that is qualitatively measured.

The community detection problem is often considered as a partitioning problem, where each node should be sorted in a meaningful community.
In this article, we propose an alternate problem: finding compact communities, and accepting that some nodes will not be sorted.
We call this sub-problem \textit{core community detection}.
To give a solution to this problem we had to consider an alternate definition of a community, where it is not the internal/external edge ratio that is measured but the internal structure.
We consider that a community can be recognized on its own, without examining neighbors, by measuring internal path length and the volume of internal edges.

Finding communities is a difficult computational issue due to the size of nowadays instances.
Big data is very common nowadays, especially in social networks.
For instance, Facebook  claims to have 1.3 billion active users, each one being friend with 130 people in average.
In this context, quadratic and super-quadratic algorithms can not be considered.

The main contributions of this article are:
1) an efficient clustering algorithm that is based on the LexDFS graph
traversal (Sec.~\ref{sec:algo}) 2) a quality function that gives a high
score to compact structures (Sec.~\ref{sec:quality}) and 3) experiments showing the practical difference between the techniques previously presented and the standard ones (Sec.~\ref{sec:experiments}).

\section{Related work}

In~\cite{leskovec_empirical_2010}, Leskovec \textit{et al.} give a methodology of community detection algorithms comparison.
They suggest that a low diameter for a large cluster is an indication of tightly-connected nodes, a community.
They compare two algorithms, and find out that the one giving lower diameter communities does not prevail when comparing conductance.

Hansen and Jaumard~\cite{hansen_minimum_1987} study the problem of cutting the network in two clusters while minimising the sum of their diameters.
They design an algorithm that performs in a cubic complexity on a complete graph.
This approach is thus not applicable to the case of real-life communities due to its high time complexity and the scaling issues that are likely to occur.


The CFinder program developped by Adamcsek \textit{et al.}~\cite{adamcsek_cfinder_2006}, one of the major overlapping community detection softwares, uses another definition of a community.
They consider that a community is a set of adjacent $k$-cliques.
We note that, as with our solution, this definition does not take into account the neighbourhood of the community.
However, they don't define a quality function (a set of nodes is a community or is not) and the time complexity of the algorithm is not satisfactory due to the need of finding all $k$-cliques of a graph.

The traditional viewpoint of partitioning the network into meaningful communities has already been questionned by Zhao \textit{et al.}~\cite{zhao_community_2011}.
They extract the local community from a node instead of considering the network as a whole.
However, they do not consider the concept of compactness and prefer to compare the in- and out-densities of the communities.

Some communtiy detection algorithms have now become standard.
The Girvan and Newman algorithm~\cite{girvan_community_2002} is the first widly recognized algorithm that solved the community detection problem.
The same ones introduced two years later~\cite{newman_finding_2004} an algorithm optimizing greedily a measure called modularity.
This algorithm were then adapted for low density graphs by Clauset \textit{et al.}~\cite{clauset_finding_2004}.
We also note that many clustering algorithms may be used for graph community detection, such as~\cite{kannan_clusterings_2004}.
For further reading, Fortunato~\cite{fortunato_community_2010} made a very thorough summary of the state of community detection in research.

The k-core decomposition, introduced by Seidman~\cite{seidman_network_1983} is close to the philosophy of what we are proposing.
A k-core is a set of vertices in which each vertex has at least $k$ neighbors.
Decomposing the network in k-cores extracts subsets with guaranteed minimum connectivity, which can be considered as the core of communities.
In a similar fashion, Wang \textit{et al.}~\cite{wang_detecting_2011} proposed to detect kernels (each vertex in the kernel has more connections inside the kernel than any vertex outside the kernel).
However, these approaches define a core by its connectivity.
Our approach is to consider the core as a well-organised structure, in which communication is efficient (\textit{i.e.} paths are short).

\section{LexDFS-based clustering algorithm}
\label{sec:algo}

We consider a set of users that have reciprocal relationships between them.
We represent them as $G=(V,E)$, an undirected graph that is formed by the set of vertices (or nodes) $V$ and edges $E$.
No self-loop is allowed.
The number of vertices is $n=|V|$, the number of edges $m=|E|$.
The number of edges incident to a node $i$ is $k_i = |\{(u,v)\in E, u=i \text{ or } v=i\}|$.
We assume that nodes and edges can have a bounded number of attributes, and we note $x.a$ the way to access the attribute $a$ of the vertex or edge $x$.
A clustering $C$ is a set of clusters, and is a partition set of $V$.
The volume of a cluster is defined as $Vol(C) = \sum_{i\in C} k_i$.
We define a satellite node $v\in V$ such that $k_v = 1$ and a connection node as a vertex which removal disconnects the graph.

\subsection{The algorithm}

This section presents our efficient (sub-quadratic) clustering algorithm that has the property of detecting the core of the communities.
It is based on the LexDFS (Lexicographical Depth-First Search) algorithm introduced by Corneil and Krueger \cite{corneil_unified_2008}, a variation of the standard DFS algorithm.
The difference lies in the choice of the next node to visit at each step.
In the standard DFS, a node is chosen uniformly at random among the neighbors of the current node.
If all the neighbors of the current node have already been visited, the neighbors of the previous node are considered, and so on.

The LexDFs algorithm makes less random choices.
Each node starts with a blank label.
When visiting the (chronologically) $i$th node, the label ``i'' is added at the start of the label of all its neighbors that have not been visited.
The node with the label that has the higher lexicographical order is chosen to be visited next.
The related pseudo-code is presented in Alg.~\ref{alg:LexDFS}.
It uses two attributes on the nodes.
The $lex$ attribute is a vector of labels used to determine the priority of neighbors. 
The $visited$ attribute marks the iteration at which the node has been visited (if it has not, it will be zero).

The LexDFS algorithm has not been very studied in the literature, except recently for theoretical research~\cite{corneil_ldfs_2013} to certify co-comparability orderings.

\begin{algorithm}
  \begin{algorithmic}[1]
    \Procedure{LexDFS}{$G, start$}
    \Require Graph $G$, Starting node $start$
    \For{$v\in V$} \Comment{Inits the attributes for every node}
    \State v.lex = ()
    \State v.visited = 0
    \EndFor
    \State stack = $\emptyset$
    \State push(stack, start)
    \State i=1
    \While{notEmpty(stack)}
    \State node = pop(stack)
    \State node.visited = i\Comment{Marks the node as visited}
    \State array = $\emptyset$
    \For{v $\in$ neighbors(node)}
    \If{v.visited = 0}
    \State remove(stack, v)
    \State v.lex = (i, v.lex)\Comment{Appends i to the label}
    \State push(array, v)
    \EndIf
    \EndFor
    \State sort(array)\Comment{Sorts by lexicographical order of the labels. Randomizes the choice between equal labels}
    \State push(stack, array)\Comment{Sets the nodes of the array on top}
    \State i = i+1
    \EndWhile
    \EndProcedure
  \end{algorithmic}
  \caption{LexDFS}
  \label{alg:LexDFS}
\end{algorithm}

We propose to use LexDFS as a basis for a clustering algorithm.
Indeed, the ordering induced by graph search has interesting
properties. Once in a community, the traversal often stays inside
since a large number of edges is intra-communities. Note that this
property is shared with the standard DFS and is closely related to the
random walk properties of the communities.  Discovering a node
increases the lexicographical order of its neighbors, which are mostly
inside the same community. They are therefore likely to be among the
next vertices to be visited.

These properties imply that nodes inside of a community are visited in a short time-lapse.
We consider the time here to be the iteration of the LexDFS.
A score may be computed for each edge, measuring the absolute time difference of the visit of the connected nodes.
\begin{equation}
  \forall e=(u,v)\in E, score(e) = 1 - \frac{|u.visited - v.visited|}{m}
\end{equation}

\begin{figure}
  \centering
  \begin{subfigure}{0.3\columnwidth}
    \includegraphics[width=\columnwidth]{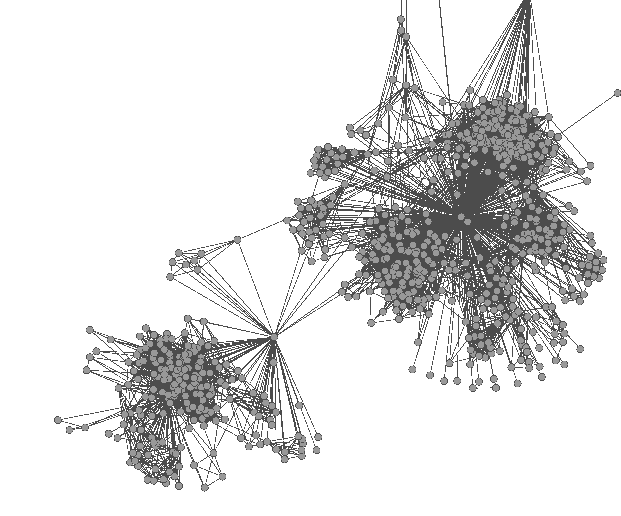}
    \caption{Full graph excerpt}
    \label{img:subfull}
  \end{subfigure}
  \begin{subfigure}{0.3\columnwidth}
    \includegraphics[width=\columnwidth]{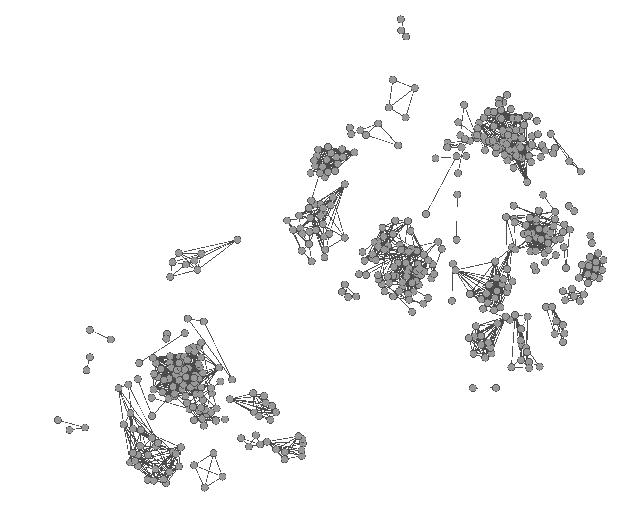}
    \caption{1800th step}
    \label{img:sub1800}
  \end{subfigure}
  \begin{subfigure}{0.3\columnwidth}
    \includegraphics[width=\columnwidth]{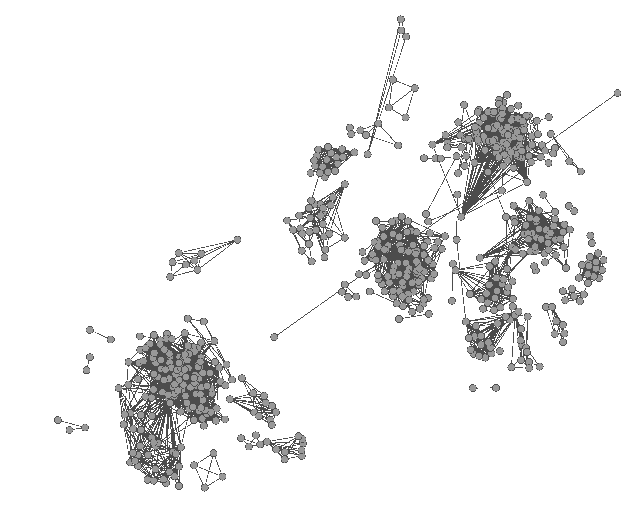}
    \caption{2700th step}
    \label{img:sub2700}
  \end{subfigure}
  \caption{State of the clusters at different iterations}
  \label{img:lexdfs}
\end{figure}

We take the mean value of this score over a few runs.
Experiments show that this mean score, after a few ($\sim 10$) runs of the LexDFS offers good topological information: filtering out the lower score edges unravels the community structure (\textit{e.g} Fig.~\ref{img:lexdfs}).
The correlation between the number of LexDFS runs and the relevance of the result is experimentally studied in Sec.~\ref{sec:experiments}.
This property is then used in the agglomerative (bottom-up) hierarchical Alg.~\ref{alg:main}.

\begin{algorithm}
  \begin{algorithmic}
    \For{$i\in [1..l]$}
    \State LexDFS(G, randnode(G)) \Comment{Starts a LexDFS on a random node}
    \For{$(u,v)\in E$} \Comment{Updates the mean value of the scores}
    \State s = 1 - |u.visited-v.visited|/m
    \State e.score = (e.score*(i-1)+s)/i
    \EndFor
    \EndFor
    \State orderedSet = E
    \State sort(orderedSet)\Comment{Sorts the edges by decreasing score}
    \State C = V
    \While{|orderedSet| $>$ 1}
    \State edge = ($v_1$, $v_2$) = pop(orderedSet) \Comment{Gets the current top-score edge}
    \State $c_1$ = cluster($v_1$) \Comment{Gets the cluster of $v_1$}
    \State $c_2$ = cluster($v_2$)
    \If{$c_1 \ne c_2$}
    \State merge($c_1$, $c_2$)
    \EndIf
    \EndWhile
  \end{algorithmic}
  \caption{Hierarchical clustering}
  \label{alg:main}
\end{algorithm}

Note that there are two random choices in the algorithm: the starting node and the choice of the next node between equal labels.
Having a deterministic choice at these decision points induces a bias.
For instance, choosing constantly a satellite node as starting point would create a large score on its edge, while they should have a low score in practice.
In the same way, choosing constantly the same edge after visiting a node would induce artificial high scores for these edges.

A run of this algorithm with 10 LexDFS iterations is presented Fig.~\ref{img:lexdfs}.
The clusters are the connected components (edges are shown if and only if they are inside a cluster) and singleton clusters are hidden.
The spatialization used is the algorithm presented by Hu in~\cite{hu_efficient_2005}.
The graph is an excerpt of the Facebook ego network presented in Sec.~\ref{sec:experiments}.
We observe in Fig.~\ref{img:sub1800} that a local community structure appears.
Interestingly, communities span and grow separately until they connect to each other.
Fig.~\ref{img:sub2700} shows that the two closely connected communities at the bottom left merge into one.

\subsection{Complexity analysis}

The algorithm time complexity is in $O(m\times
log(m)+(m+n)\times l)$, where $l$ is the number runs of the LexDFS.

\subsubsection{LexDFS}

LexDFS visits each node once and each edge twice.
Visiting an edge (l.15-17 of Alg.~\ref{alg:LexDFS}) is in $O(1)$ with the appropriate data structure.
Indeed, if the vertex links to its position in the stack, the removal can be done in constant time.
Adding a label to a vertex should also be linear if a linked list is used to represent labels.
We consider that the array of neighbors is of size $n-1$, which is the maximum number of neighbors.
A variable keeps track on the actual size of the array.
This data structure enables constant time addition of a vertex and standard sorting. 

However, the complexity of the sorting operation (l.20) is not immediate, since the comparison is not in $O(1)$, but depends on the size of the label of the elements.
We examine the case of a vertex which has $d$ neighbors.
We note that the $ith$ neighbor in the size $d$ array has a degree $k_i$ and, in the worst-case scenario, it has already been discovered by all its neighbors.
Therefore, the label has $k_i$ elements and comparing the label of the $ith$ and $jth$ node takes $O(min(k_i, k_j))$.

In standard sorting algorithm such as quick sort or merge sort, two different elements are never compared twice.
Taking $m$ such as $k_m = max_i(k_i)$, we therefore have an upper bound of the cost of the comparisons (with $\overline{k}$ being the mean degree of the neighbors):
\begin{align*}
  cost & < & \sum_{i=1}^d\sum_{j=i+1}^dmin(k_i, k_j) & < & d\times \sum_{i=1}^dmin(k_i, k_m) \\
    & & & = & d^2\times \overline{k}
\end{align*}

This cost is summed over all the vertices, the total cost of the comparisons is thus in $O(\sum_{v\in V} k_v^2\overline{k_v})$.
We assume that the distribution of the degree of a neighbor of a uniformly selected vertex is seemingly the same as the uniform degree distribution, therefore $O(\overline{k}) = O(d)$.
Since $\forall (a,b)\in \mathbb{R}^{+2}, a^2+b^2\le (a+b)^2$, the total cost of lexicographic comparisons is in $O(n\times \overline{d}^3)$, where $\overline{d}$ is the mean degree.

The complexity of the LexDFS algorithm is therefore in $O(m + n\times \overline{d}^3)$.
In social networks, the degree distribution is considered scale-free : its probability distribution is not affected by the size of the network.
Therefore the mean degree may be considered constant, \textit{i.e.} $O(d) = O(1)$.
The final complexity of the LexDFS algorithm is thus in $O(n+m)$.

\subsubsection{Cluster merging}

Computing the score of the edges is in $O(m)$.
Sorting the edges given their score is in $O(m\times log(m))$.

The successive merges of the clusters is a case of union of disjoint sets.
This problem has been solved by a quasi-linear algorithm by Tarjan in \cite{tarjan_efficiency_1975}, in in $O(m \times \alpha(m))$.
Since $\alpha(m)$ is the inverse of the Ackerman function, it is
\textit{very} slowly growing. Since it grows slower than a logarithmic function, the complexity of the merging is dominated by $O(m\times log(m))$.

The complexity of the whole algorithm is therefore in $O(m\times log(m)+(m+n)\times l)$.

\section{Quality functions}
\label{sec:quality}

A quality function $f:\mathcal{C} \rightarrow \mathds{R}$ (where $\mathcal{C}$ is the set of all possible clusterings) may be defined to set a value evaluating clusterings.
Its more immediate application is the comparison of the result of clustering algorithms.
For a hierarchical clustering algorithm, a quality function is even more crucial: since each step produces a distinct clustering, the quality function gives us indications on which one are the best.
In this case, the quality function helps selecting relevant solutions.

\subsection{Conductance}
\label{subsec:conductance}

Kannan \textit{et al.}~\cite{kannan_clusterings_2004} tried to find cuts that were meaningful for clustering.
Instead of cutting the minimum number of edges, their cuts were minimal with respect to a quality function.
They defined the \textit{conductance} of a cluster, corresponding to the probability that a random walk will exit the cluster.
In practice, it measures the external degree of the cluster over the volume of the cluster.
The conductance of a clustering is defined as the minimum conductance of the clusters composing the clustering (low values indicate here a high quality).
\begin{equation}
  \label{eq:conductance}
  \phi (C) = min_{c\in C} \frac{|\{(u,v)\in E, u\in c \text{ and } v\notin c\}|}{min(Vol(c), 2m - Vol(c))}
\end{equation}

This quality function is unsatisfactory since the aggregation method may lead to misleading comparisons between communities.
For instance, a graph that includes a satellite node would maximize conductance just by having one cluster with just the satellite and another for the rest of the graph.
On the other hand, even if it is not adapted to the clustering level, this measure shows meaningful results when applied on individual clusters (see experiments Sec.~\ref{sec:experiments}).

\subsection{Modularity}

The most popular quality function in the last decade \cite{brandes_modularity_2008}, \cite{clauset_finding_2004}, \cite{fortunato_resolution_2007} is the \textit{modularity}, $Q$ defined by Newman in \cite{newman_finding_2004} as:
\begin{equation}
  \label{eq:modularity}
  Q(C) = \sum_{c\in C} \left[ \frac{E(c)}{m} - \left( \frac{Vol(c)}{2m} \right)^2 \right]
\end{equation}
Where $E(c)$ is the number of edges connecting the vertices of the nodes inside the cluster $c$.

The first part of the sum is called \textit{coverage}.
It represents the fraction of edges inside the clusters.
The second part is, for each cluster, the expected value of the coverage when applying the configuration model to the full graph.
It is a simple model that, given a degree distribution, connects every half edge (or ``stub'') to another with uniform probability.
Modularity therefore detects if a group of nodes is unexpectedly tied together, but does not assume any internal structure.
This perception of a community agrees with the standard definition presented in Sec.~\ref{sec:intro}.

\subsection{Compactness}

We define here our quality function, the compactness.
Modularity gives importance in a community to the internal number of edges compared to the external one.
Even disregarding the issue of modularity with scaling~\cite{fortunato_resolution_2007}, we believe that it misses an important point: the shape of the community.

\begin{figure}
  \centering
  \includegraphics[width=0.25\textwidth]{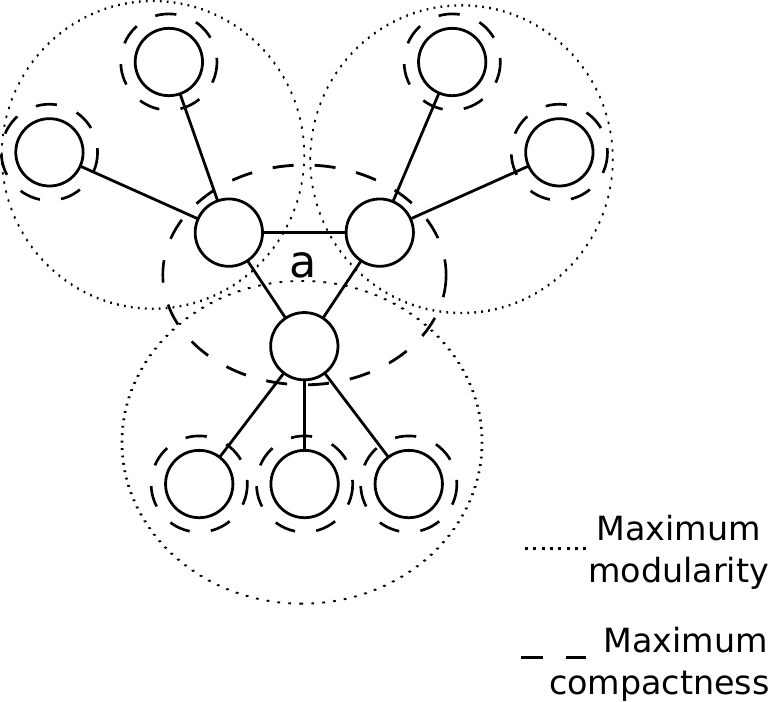}
  \caption{A simple case where the standard definition of a community is not satisfactory}
  \label{img:pathological-case}
\end{figure}

In Fig.~\ref{img:pathological-case}, the central 3-clique $a$ is not a community in the weak or the strong sense.
Nevertheless, a human would identify the central 3-clique as a single entity, and would therefore not agree with the standard definition of a community.
This example works exactly the same if the satellite nodes are any other kind of subgraph, other 3-cliques for instance.
Even more convincing is the generalisation to a central n-clique, where each node is connected to $n$ different satellites.

What makes communities in Fig.~\ref{img:pathological-case} stand out is their internal structure.
Since every node is connected to another, communication between individuals is instantaneous.
But, in a general case, an information with general purpose passes through intermediates to reach its destination.
We model such an information transfer as a perfect broadcast communication process, that is the information reaches the individual $i$ at a time $t$ iff the information reached one of the neighbors of $i$ at time $t-1$.
The efficiency of a connected subgraph regarding such an information transfer model can be quantified as the average transfer rate before the stable state.
A lot of people reached in a little time implies a good quality of the subgraph, and a clique is the best subgraph in that case.
This definition is very intuitive : a characteristic of a group of friends or a family is that important news reach everyone very quickly.

However, this definition makes the very strong assumption on the underlying transfer model that the communication is perfect.
It is not true in most cases of observed communications, but a strong correlation between the number of neighbors that have transferred the message and the probability of transfer is common knowledge.
We therefore take into account the density of the subgraph, and we use the average edge rate for the process instead of the average node rate to quantify quality.
We define a \textit{compact community} as subsets of vertices within which vertex-vertex connections are dense, but the length of paths is small.
The length of paths may be represented by the mean eccentricity (the expected average edge rate for a random source) or the diameter (the edge rate for the worse source).
On this basis, we define an alternate quality measure, the \textit{compactness} ($L$).
\begin{eqnarray*}
  L(C) & = & \sum_{c\in C} L(c) \\
  L(c) & = & \left\{
  \begin{array}{l l}
    0 & \quad \text{if $E(c)=0$}\\
    \dfrac{E(c)}{diam(c)} & \quad \text{otherwise}
  \end{array} \right.
\end{eqnarray*}

where $diam(c)$ is the diameter of the sub-graph induced by the cluster $c$.
The compactness of a cluster that has no edge (and therefore a zero diameter) is zero as well.
Note that the measure can be simply normalized with a division by the total number of edges.

This quality function does not always favor the communities as defined in the introduction, weak or strong.
The (normalized) measure is also close to the coverage, with the added property that the internal organization of the cluster is taken into account.
A disorganized, spread cluster has a low quality, while a compact one is considered as high quality.

The maximum compactness of a graph with $n$ vertices is attained by a n-clique, and therefore the better clustering of a n-clique consists of an only cluster containing the whole graph.
On the other hand, when a community structure is visible, experiments showed that the whole graph is not the optimum clustering.
Much higher values may be attained by regrouping the well-connected core of communities.

\subsection{Comparison between quality functions}

We now compare compactness and modularity, showing the different choices they make in examples.
We will not include conductance here, since it is very similar to modularity in all the chosen examples.

\begin{figure}
  \centering
  \begin{subfigure}{.6\columnwidth}
    \includegraphics[width=\columnwidth]{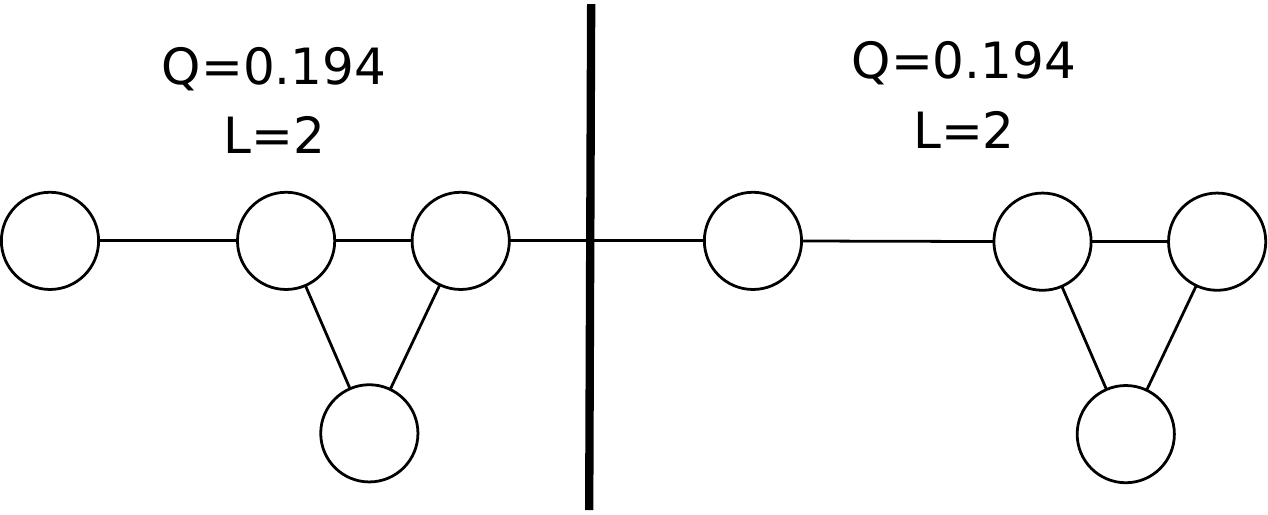}
    \caption{Clustering including satellite and connection nodes}
    \label{img:quality2}
  \end{subfigure}
  \begin{subfigure}{.6\columnwidth}
    \includegraphics[width=\columnwidth]{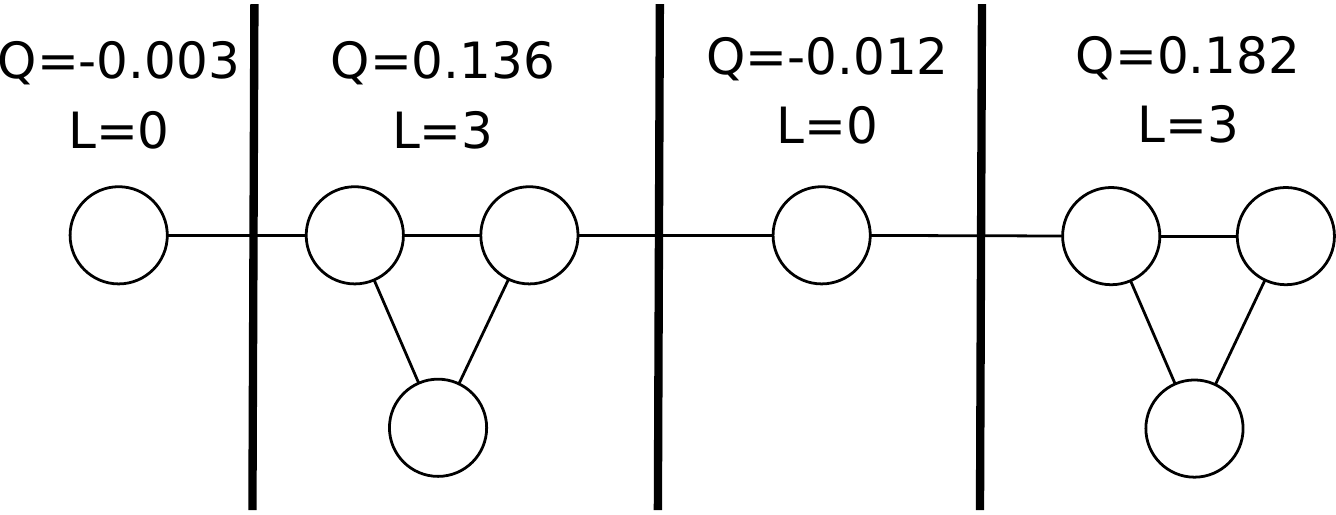}
    \caption{Clustering excluding satellite and connection nodes}
    \label{img:quality1}
  \end{subfigure}
  \caption{Example : modularity (Q) and compactness (L)}
  \label{img:quality-example}
\end{figure}

Modularity has the tendency to give a higher score to the largest possible meaningful clusters, including even satellite and connection nodes, \textit{e.g.} Fig.~\ref{img:quality-example}.
On this example, the clustering presented Fig.~\ref{img:quality2} has a better total modularity score than Fig.~\ref{img:quality1}, classifying the satellite node and the connection node.

It may be considered sound from a classification perspective of the community detection problem (every node has to be classified).
However, this quality measure is not adapted when the aim is to find the connected center of communities, and to leave the rest.
The latter approach is in some sense more intuitive: when a user is only tied to one person belonging to a community, it may not be relevant to classify him in this community.\\
Another advantage of compactness against modularity is presented Fig.~\ref{img:pathological-case}.
Since modularity follows the standard definition of a community, unrealistic situations might arise where a clique is split between clusters.
Compactness, on the other hand, detects compact clusters which is in some situations more realistic.

We can also compare these measures in terms of locality.
How does a modification of the community (addition/deletion of an edge/node) change the quality score?
We differentiate \textit{local} methods, that need only to recompute values in the neighbourhood of the change, and \textit{global} methods that need to recompute everything.
For conductance, the out-degree and the volume can be locally computed.
Modularity needs the same information, it is once again a local computation.
On the other hand, compactness is a global computation since even a far away node can have an important impact on the diameter.
The choice to take the diameter over the mean distance means that one node being added to a big cluster might greatly change the structure.
Diameter might suddenly rise by adding satellite nodes, or decrease by adding a central node to the cluster.
Therefore, a bad structure is more "punished" by a bad score when taking the diameter.
On the other hand, the mean distance is a more stable and a more
easily approximated value.\\
What makes the strength of compactness is also its main drawback.
Having to recompute the diameter every time a node is added induces a huge time complexity.
Modularity, on the contrary, may be computed on the fly with almost no additional cost.
However, it is not problematic since it may be used only to rate a clustering algorithm, and therefore it is only needed for test purposes.

\subsection{Compliance to quality function axioms}

Van Laarhoven \textit{et. al.} produced a set of axioms that a quality function should intuitively use in a context of graph clustering~\cite{van_laarhoven_axioms_2014}.
No unparametered quality function complying to all these axioms is presented in their work, and they proved that modularity does not comply to two of them.
We prove here that compactness complies to all of these axioms.
We note that these proofs could be applied to a mean distance alternative of the compactness with little to no modification.

First, we need to extend our definition to weighted graphs.
We consider that weights correspond to communication speed.
The distance between two points is the time needed to go from one point to another.

We comply to the notation used in the main article, stating that weights are positives and a zero weight corresponds to the non-existence of an edge.
Because we use a diameter, we need a distance function between nodes with weighted edges.
\begin{eqnarray*}
dist(u, v) & = & min_{\pi=(v_1...v_k)\in V^k}(\sum_{v_i, v_{i+1}\in \pi} 1/w(v_i, v_{i+1})) \\
L(c) & = & \sum_{e\in E(c)} w(e) / max_{u,v\in C}(dist(u,v)) \\
\end{eqnarray*}
We now check the axioms one by one.

\subsection{Permutation, scale invariance, continuity}
\noindent{\bf Permutation}. Trivial since only the information on the clustering and the neighborhood of nodes is used.

\noindent{\bf Scale invariance}.
Let $G'=(V, E')$ and $\alpha \in \mathbb{R}_{>0}$ such that $E'=\{e'=(u,v), \exists e=(u,v)\in E, w(e') = \alpha w(e)\}$.
We have $L(c) = E(c)/diam(c)$.
$diam(c') = diam(c)/\alpha$ because the minimum path between two nodes does not change if the edges are just modified by a multiplicative positive constant.
$L(c') = \alpha^2 E(c)/diam(c) = \alpha^2 L(c)$
Therefore, $\forall C_1, C_2\in C$, if $L(G, C_1) \ge L(G, C_2)$ then $\alpha^2L(G, C_1) \ge \alpha^2L(G, C_2)$, therefore $L(G', C_1) \ge L(G', C_2)$.

\noindent{\bf Continuity}. infinitely small modification of the weight of the edges leads to infinitely small modification of the quality.

\subsection{Locality}
Since compactness is not influenced by intra-cluster edges, we have
that $\forall C1, C2\in \mathcal{P}(V), C1\cap C2 = \emptyset, L(G, C1 \cup C2) = L(G, C1) + L(G1, C2)$.
Therefore,
\begin{eqnarray*}
  & & L(G1, C_a \cup C_1) \ge L(G1, D_a\cup C_1) \\
  & \leftrightarrow & L(G1, C_a) \ge L(G1, D_a) \\
  & \leftrightarrow & L(G2, C_a\cup C_2) \ge L(G2, D_a\cup C_2)
\end{eqnarray*}

\subsection{Richness}
Let G be a clique graph created from a clustering C (two vertices are connected iff they are in the same cluster in C).
Let D be a clustering of maximal compactness.
If two unconnected points of G are in different clusters in D, the quality of the cluster in D is zero (infinite distance).
Splitting the unconnected points in two clusters results in a better or equal clustering.
If all points of two clusters in D are connected, the cluster
resulting of the fusion is a better quality cluster.
Hence, C is a maximum compactness clustering on G.
C is arbitrary, thus any clustering is a maximum-compactness clustering on the corresponding clique graph.

\subsection{Monotonicity}
Decreasing the weights of intra-cluster edges does not change the quality value.
We thus focus on the increase of inter-cluster edges.
If the weight of such an edge is increased, it was either on the unique maximum shortest-path or not.
In the first case, the distance of the path has decreased, therefore the quality has increased.
In the second case, it has either changed to a shortest path between the two nodes at maximum distance (and the quality has increased) or it has not and the quality has not changed.
A consistent improvement thus implies an equal or increased compactness, which proves monotonicity.

\section{Experiments}
\label{sec:experiments}

For reproducibility, the code and the results of the experiments are available at \verb?creusefond.users.greyc.fr?.

Experiments were carried out on an intel i5-3670 (3.4 GHz) CPU with 8GB of RAM.
The convergence tests were computed on an AMD Opteron 6174 2,2 GHz CPU with 256GB of RAM.
Our clustering algorithm runs in a few seconds, even for the largest graphs.

We used three real-world graphs for tests (taken from
SNAP\footnote{http://snap.stanford.edu/data/}): 
Facebook~\cite{leskovec_learning_2012} ($n = 4039$, $m = 88234$),
astro~\cite{leskovec_graph_2007} ($n = 18772$, $m = 198110$) and enron~\cite{klimt_introducing_2004} ($n = 36692$, $m = 183831$).

The first algorithm to be used is of course the LexDFS-based algorithm presented in this article.
Unless stated otherwise, we make it run for 20 iterations, which seems sufficient considering the variation of compactness along multiple runs.

We compare our method with a modularity optimisation algorithm, introduced by Newman \textit{et al.}~\cite{newman_finding_2004} and modified by Clauset \textit{et al.}~\cite{clauset_finding_2004} for low density graphs.
It is a hierarchical algorithm that starts with a community per node and then merges the communities that bring the best increase in modularity.
The algorithm from Clauset \textit{et al.} is loglinear, as ours.

\subsection{Description of the experiments}

\subsubsection{Single clusters comparison}

This experiment aims to compare the quality of single clusters instead of a clustering as a whole.
This method, introduced by Leskovec \textit{et al.}~\cite{leskovec_statistical_2008} with their NCP (Network Community Profile) plots, is simply to plot the quality function related to the size of the cluster.

Indeed, one can notice that every quality function presented here is described as the agglomeration of a function applied to single clusters.
In this experiment, we are only interested in this function and the result on single clusters.
Note that a low conductance indicates a high quality, unlike the other quality functions.

\subsubsection{Global clustering quality}

By definition, at every step of a hierarchical algorithm is associated a clustering.
If the application requires to detect the different communities of a network, we will be interested in the evolution of the score of a quality function during the execution of the algorithm.
The global maximum score represents the best clustering according to the quality function and local variations might tell us important information about the behavior of the algorithm.

We compare the modularity and compacity for the two considered algorithms.
We don't show conductance since it is not suited for a global quality measurement (see Sec.~\ref{subsec:conductance}).
In order to have an idea of the possible variations induced by the random choices made by our algorithm, we ran the experiment 20 times and showed minimum, maximum and mean values for each iteration.

\subsubsection{Convergence}

In order to have an idea of the relation between the convergence of the scores and the number of times a LexDFS is started, we measure at each LexDFS computation the difference of ordering of the edges.
Indeed, the output ordering induces how the hierarchical clustering behaves.
We note $\forall i\in \{1..l\}$, $o_i:E \rightarrow \{1..m\}$ a set of functions indicating the order of an edge at the $i$th iteration.

We first compute the number of edges that have a different spot in this ordering.
But this seems very restrictive since the precise order is sometimes not meaningful and does not change the final cluster. 
The most important here is that the edges stay in the same area of their order.

We define $w$ as a window, or an error tolerance.
We compute $c_i(w)$, the number of edges that are not in the window, and consider that the system converges if it goes to zero. 
\begin{equation}
  c_i(w) = m - |\{e\in E, o_i(e) - w \le o_{i+1}(e) \le o_i(e) + w\}|
\end{equation}

We are interested in a size-one window, which means that the order of the edges must be equal.
We also considered a window of arbitrary size, here 20.
Finally, we also considered a window proportional to the number of edges, here $2w\approx 0.01m$ (the 1\% window).

\subsection{Results}

\subsubsection{Single clusters comparison}

\begin{figure}
  \centering
  \begin{subfigure}{0.4\columnwidth}
    \includegraphics[width=\columnwidth]{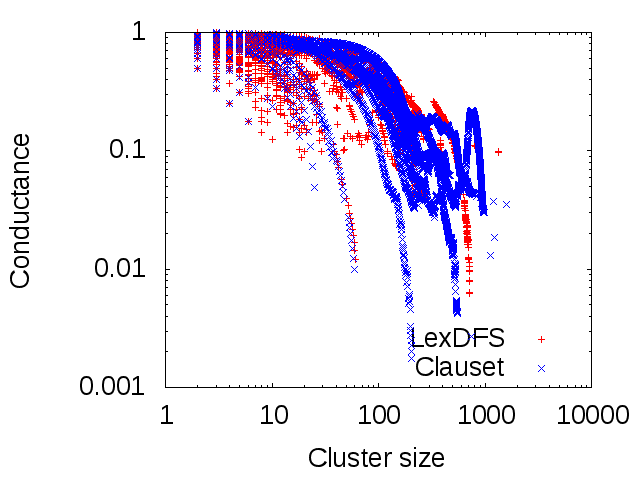}
    \caption{Facebook dataset}
    \label{img:local_cond_face}
  \end{subfigure}
  \begin{subfigure}{0.4\columnwidth}
    \includegraphics[width=\columnwidth]{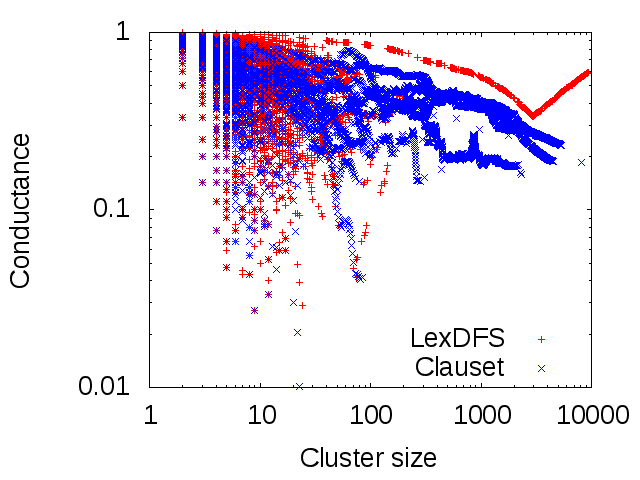}
    \caption{Astro dataset}
    \label{img:local_cond_astro}
  \end{subfigure}
  \begin{subfigure}{0.4\columnwidth}
    \includegraphics[width=\columnwidth]{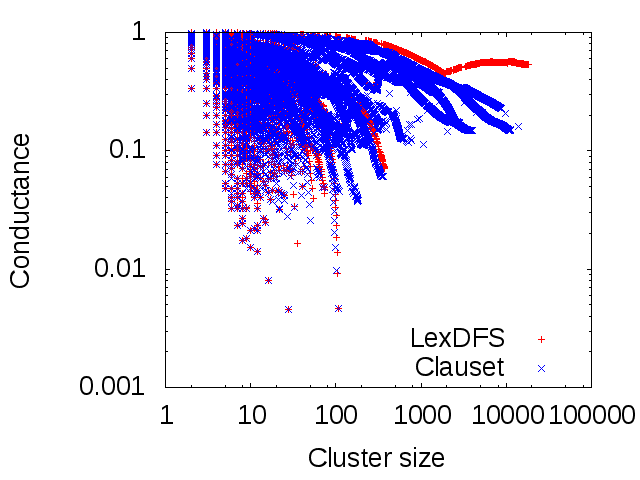}
    \caption{Enron dataset}
    \label{img:local_cond_enron}
  \end{subfigure}
  \caption{Cluster conductance comparison}
  \label{img:local_cond}
\end{figure}

The conductance plot Fig.~\ref{img:local_cond} shows that the
modularity optimisation algorithm is better at optimizing conductance.
Conductance on small clusters is approximately the same, but the
LexDFS-based algorithm does not perform well on large clusters
regarding conductance.  It is not particularly visible on the smallest
dataset (Fig.~\ref{img:local_cond_face}).  Both methods reach low
conductance clusters, small ($\sim$ 80 nodes) and large ($\sim$ 800 nodes).

Fig.~\ref{img:local_cond_astro} and Fig.~\ref{img:local_cond_enron} shows that large clusters ($>$300 nodes) with low conductance($<$0.3) are created by the modularity optimisation
method.  Our method does not feature such cluster, but has a similar
performance on small clusters.  We can also see from these plots that
the lexdfs-based algorithm has the tendency to form a giant cluster
quite quickly.

\begin{figure}
  \centering
  \begin{subfigure}{0.4\columnwidth}
    \includegraphics[width=\columnwidth]{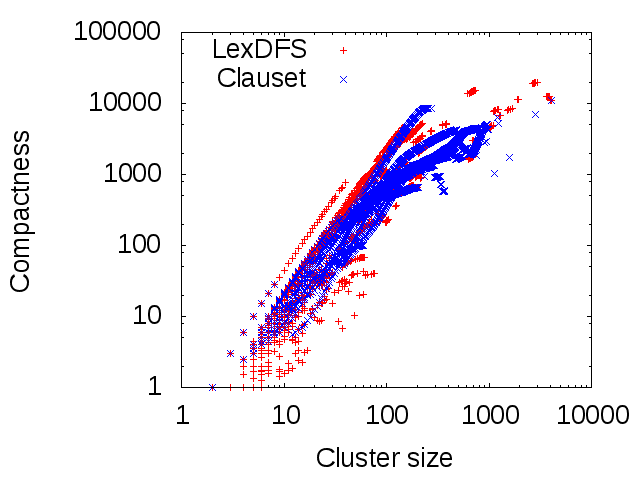}
    \caption{Facebook dataset}
    \label{img:local_comp_face}
  \end{subfigure}
  \begin{subfigure}{0.4\columnwidth}
    \includegraphics[width=\columnwidth]{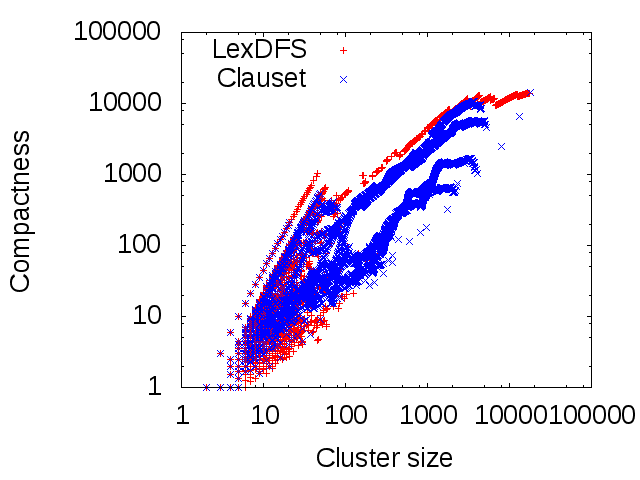}
    \caption{Astro dataset}
    \label{img:local_comp_astro}
  \end{subfigure}
  \begin{subfigure}{0.4\columnwidth}
    \includegraphics[width=\columnwidth]{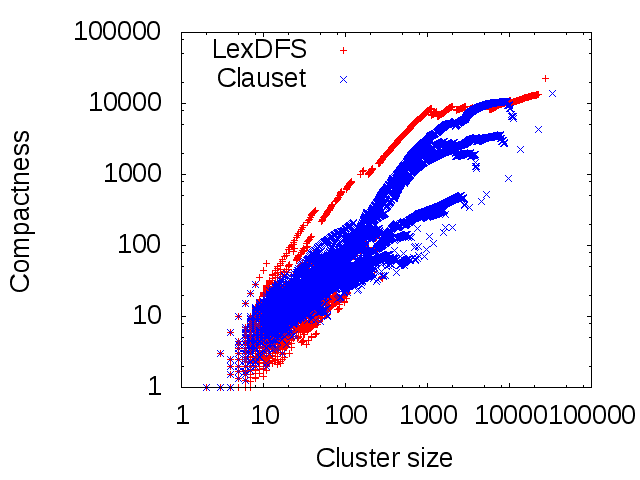}
    \caption{Enron dataset}
    \label{img:local_comp_enron}
  \end{subfigure}
  \caption{Cluster compactness comparison}
  \label{img:local_comp}
\end{figure}

The compactness plots Fig.~\ref{img:local_comp} show, on the other hand, better results for our algorithm.
Fig.~\ref{img:local_comp_face} shows that our method produces clusters of higher compactness than the ones produced by the modularity optimisation method.
The largest clusters have an especially high compactness, and our methods provide many more large clusters ($>$1000 nodes) than the other one.

This experiment applied to the other datasets (Fig.~\ref{img:local_comp_astro} and ~\ref{img:local_comp_enron}) give slightly different results.
Both methods still produce similar small clusters ($<$100 nodes).
On the other hand, the unique large cluster ($>$300 nodes) that was observed in the previous experiment has a higher compacity than the clusters produced by the modularity optimisation algorithm.

\subsubsection{Global clustering quality}

\begin{figure}
  \centering
  \includegraphics[width=.6\columnwidth]{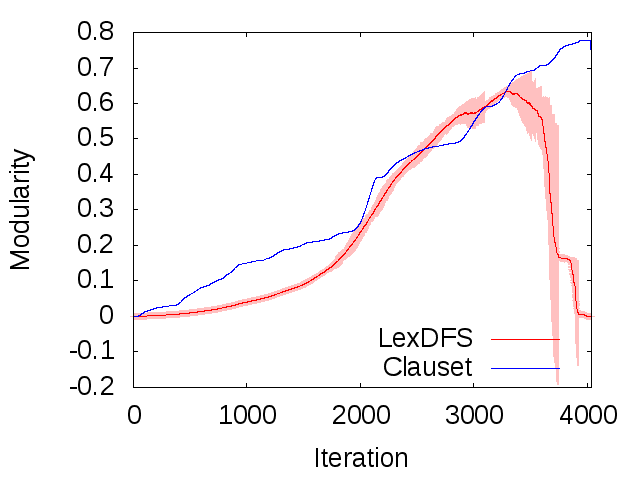}
  \caption{Clustering modularity comparison, facebook dataset}
  \label{img:global_mod}
\end{figure}

\begin{figure}
  \centering
  \begin{subfigure}{.6\columnwidth}
    \includegraphics[width=\columnwidth]{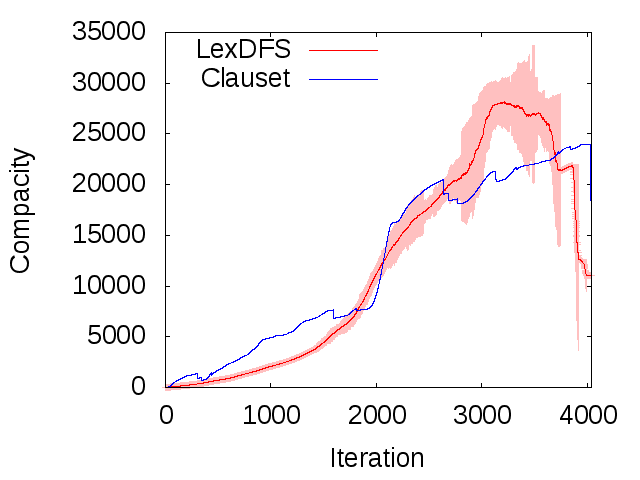}
    \caption{Facebook dataset}
    \label{img:global_comp_face}
  \end{subfigure}
  \begin{subfigure}{.6\columnwidth}
    \includegraphics[width=\columnwidth]{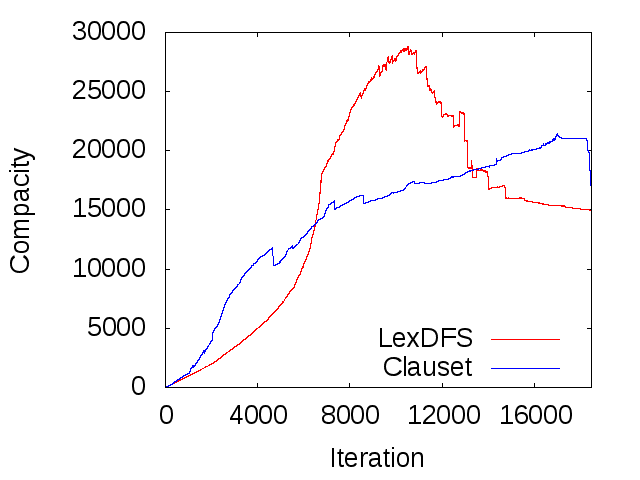}
    \caption{Astro dataset}
    \label{img:global_comp_astro}
  \end{subfigure}
  \caption{Clustering compactness comparison}
  \label{img:global_comp}
\end{figure}

As we can see in Fig.~\ref{img:global_mod}, the global modularity quality is higher in the case of the greedy algorithm.
We noticed a similar effect on other datasets, not presented here for shortness.

However, maximum compactness is reached by our algorithm on every dataset considered, see Fig.~\ref{img:global_comp}.
As we can see in Fig.~\ref{img:global_comp_face}, the clustering produced by our algorithm is less compact than its competitor at some steps.
For instance, the first thousand iterations of the greedy algorithm produce high-compactness results.
This can be explained by the different behaviour of both algorithms, our algorithm creates and enlarges multiple small clusters while the other one creates one cluster after the other.

We conclude that comparing two algorithms step-by-step is not always meaningful since they use different methods.
On the other hand, it is relevant to compare maximum clusterings and to get some insight on the quality of the successive clusterings.

\subsubsection{Convergence}

\begin{figure}
  \centering
  \includegraphics[width=0.3\textwidth]{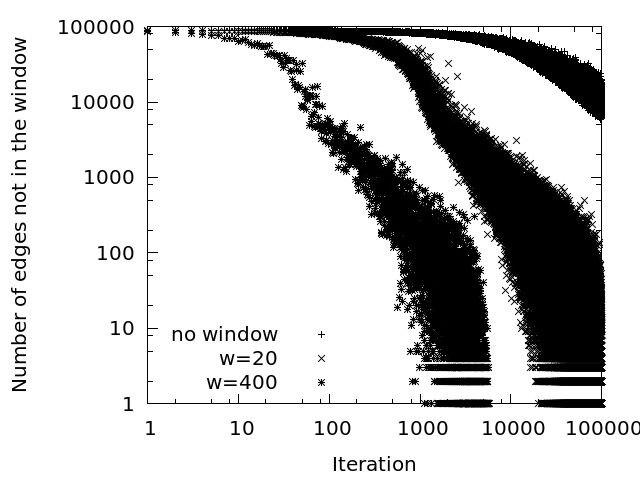}
  \caption{Convergence with different acceptance windows}
  \label{img:convergence}
\end{figure}

Fig.~\ref{img:convergence} presents the results of the convergence experiments.
It is easy to prove, as observed, that a larger window induces lower scores: if $w_1 < w_2$, the edges that are in the window $w_1$ are also in the window $w_2$, therefore $\forall i$, $c_i(w_1) \ge c_i(w_2)$.
After less than 100 iterations, only 10\% of the edges are outside of the 1\% window.
However, most of these edges are not in the closer window $w=20$.
Indeed, we observe that this 10\% boundary is reached after $\sim$ 1200 iterations.
The windowless test, as expected, gives very bad results since the 10\% boundary is not even reached after $10^5$ iterations.
This experiment could be used to have an estimate on how many LexDFS runs would be pertinent.
However, it should be treated with care: if an edge is not in the same window at different iterations, this does not mean that the resulting clustering will be different.
On top of that, the remaining edges that are not in the same window may be structurally important edges (\textit{e.g.} high betweenness centrality) and their order might be of great importance.

\section{Conclusion}

We presented an alternate definition of the community detection problem based on the idea that a community needs to be compact.
To our knowledge, this approach has not been considered before, and leads to a new quality function.
We defined an efficient community detection algorithm.
We compared it with an existing standard algorithm on real-world networks and it finds compact communities, outperforming the modularity optimisation algorithm.

\bibliographystyle{abbrv}
\bibliography{biblio}

\end{document}